\documentclass[10pt]{article}

\pdfoutput=1

\usepackage[utf8]{inputenc}
\usepackage[T1]{fontenc}
\usepackage{graphicx}
\usepackage{color}
\usepackage{listings}
\usepackage{amsmath}
\usepackage{amssymb}
\usepackage{hyperref}
\usepackage{rtsched}
\usepackage[backend=bibtex]{biblatex}
\newcommand{\R}{\ensuremath{\mathbb{R}}}
\newcommand{\N}{\ensuremath{\mathbb{N}}}

\title{Real-Time scheduling: from hard to soft real-time systems}
\author{Giuseppe Lipari \\ Université de Lille 1\\giuseppe.lipari@univ-lille1.fr \and Luigi Palopoli \\Università di Trento\\paolopoli@unitn.it}

\definecolor{mygreen}{rgb}{0,0.6,0}
\definecolor{mygray}{rgb}{0.5,0.5,0.5}
\definecolor{mymauve}{rgb}{0.58,0,0.82}

\lstdefinestyle{customc}{
  belowcaptionskip=1\baselineskip,
  breaklines=true,
  xleftmargin=\parindent,
  language=C,
  showstringspaces=false,
  basicstyle=\footnotesize\ttfamily,
  keywordstyle=\bfseries\color{mygreen},
  commentstyle=\itshape\color{mymauve},
  identifierstyle=\color{blue},
  stringstyle=\color{red},
  numbers=left,
  numberstyle=\tiny,
}

\begin{document}

\maketitle

\begin{abstract}
  Real-time systems are traditionally classified into hard real-time
  and soft real-time: in the first category we have safety critical
  real-time systems where missing a deadline can have catastrophic
  consequences, whereas in the second class we find systems for which
  we need to optimise the Quality of service provided to the user.
  
  However, the frontier between these two classes is thinner
  than one may think, and many systems that were considered as hard
  real-time in the past should now be reconsidered under a different
  light. 

  In this paper we shall first recall the fundamental notion of
  time-predictability and criticality, in order to understand where
  the real-time deadlines that we use in our theoretical models come
  from.  We shall then introduce the model of a soft real-time system
  and present one popular method for scheduling hard and soft
  real-time tasks, the resource reservation framework.

  Finally, we shall show how resource reservation techniques can be
  successfully applied to the design of classical control systems,
  thus adding robustness to the system and increasing resource
  utilisation and performance.
\end{abstract}

\section{Introduction}
\label{sec:intro}

Real-time systems are computational systems whose correctness depends
not only on the correctness of the produced results, but also on the
time at which they are produced. They must interact with their
external environment in a timely manner: for example, real-time
systems that control physical plants (digital control systems, or the
so-called \emph{cyber-physical systems}) must perform their
computation and produce their results at the typical timing rates of
the physical environment, and within a bounded delay.

Consider a brake-by-wire control system in a
modern car. It is surely a real-time system, as it must command the
brake with a maximum delay from when the driver presses the pedal,
otherwise a dangerous accident may be produced. As last
example, consider a multimedia player in an embedded system that must
reproduce a movie in a predefined regular periodic rate to provide a
high \emph{quality of service} to its users.

It is clear from these examples that there are many types of timing
constraints to be considered, depending on the definition of
\emph{correctness}, and the robustness of the system to a violation of
such timing constraints.

\subsection{Hard real-time}
\label{sec:intro-hard}

Traditionally, real-time researchers have focused on the notion of
\emph{hard real-time system}. An hard real-time system is often
modelled as a set of computational tasks to be executed concurrently
on the selected hardware platform by a \emph{real-time
  scheduler}. Computational tasks are characterised by a Worst-Case
Execution Time (WCET). They are recurrently activated by input stimuli
(i.e. external interrupts, internal events or periodic timers) with a certain
\emph{activation-pattern}. Furthermore, they are assigned a
\emph{relative deadline}, that is, a maximum delay in completing the
computation from when an input stimulus activated the task. 

The design of an hard real-time system is subject to the constraint
that all instances of all tasks must complete their computation within
their assigned relative deadline. The system is modelled with a
mathematical formalism, and we can answer several interesting
questions:
\begin{itemize}
\item \textbf{Feasibility}: does there exist a schedule in which all
  deadlines are respected?
\item \textbf{Schedulability}: given a specific scheduler, is it able
  to generate on-line a schedule such that all deadlines are
  respected?
\item \textbf{Parameters assignment}: find the \emph{optimal}
  assignment of scheduling parameters to tasks so that all deadlines
  are respected.
\end{itemize}

Several variations of the same problem have been investigated in the
real-time system literature, by varying the task model
(e.g. multi-frame tasks \cite{baruah1999generalized}, generalised
recurring branching tasks \cite{baruah1998feasibility}, DAG tasks
\cite{stigge2011digraph}, etc.), the platform model (single or
multiprocessor, homogeneous or heterogeneous), the interaction model
(independent tasks, precedence constraints \cite{chetto1990dynamic},
shared resources \cite{sha1990priority}, etc.).

However, a legitimate question is: what happens if a deadline is
missed?

\subsection{Missing deadlines}
\label{sec:intro-soft}

There are several possible outcomes of a deadline miss.  Let us fist
consider that case in which a deadline miss causes a software bug.
For example, it may happen that the program goes into an inconsistent
state (some data structure contain inconsistent data, or pointers
contain the wrong address in memory, etc.). In this case, the software
may start producing inconsistent outputs or even crash.

Basic software development guidelines tell us that this should never
happen. Even if we had previously performed a careful analysis of the
code and a precise schedulability analysis which ``guarantees'' that no
deadline miss will occur, there is always the possibility that our
mathematical model does not capture precisely all possible effects, or
that an unexpected hardware fault compromises the ability of a task to
complete before its deadline. Therefore, even for hard real-time
systems that have been deemed schedulable by off-line schedulability
analyses, it is a good design norm to make sure that the software
remains into a consistent state in case of a deadline miss so that it
can continue its execution in degraded mode. For example, if the
output of Task A is put into a buffer that is later read by another
Task B, we need to make sure that the buffer content is always valid, no
matter if the Task A completes before its deadline or not.

From now on, we will assume that the system is developed according to
good design norms, and that the software data structures remain
consistent under all operating conditions.

Having ruled out software bugs, what may happen when a deadline is
missed? the answer to this question depends on 1) the application
requirements, 2) the software architecture and how a deadline miss is
dealt with.

As a first example, consider a digital control system consisting of an
embedded board with a real-time operating system executing a set of
real-time control tasks. Suppose that Task $A$ writes the results of its
computation in a buffer that is later used by the actuators to
command the physical system. If $A$ is late, the buffer will hold the
previous value of the command at the time of the actuation, so the
actuator will use the old data to command the plant. The impact of
this on the dynamics of the physical system depends on the control
law, the dynamic evolution of the physical system, the sampling
period, etc. We can estimate that, if the deadline miss happens too
often or too many times consecutively, the control system may become
unstable to the point that we cannot control it anymore. This can lead
to serious consequences in critical systems like cars and planes. All
these consideration pertain to the theory of dynamic control, so we
can answer our original question only after analysing the overall
system, including the control dynamic.

Consider now the example of a multimedia player, where a set of tasks
is in charge of playing a movie by decoding the flow of data frames
into video and audio data to be synchronised and set to the
appropriate output video and sound peripherals. Typically, at 40 video
frames per second, one video frame must be decoded and shown on the
screen every 25 milliseconds. What if the decoding task misses its
deadline? If the system is well designed, we will experience some
strange artifacts on the screen (e.g. large blocks of pixels, low
resolutions areas, etc.) which will lower our appreciation of the
show. If this happens too often, the experience may become very
unpleasant. The exact characterisation of the \emph{Quality
  of Service} perceived by the user depends, once again, on the
requirements of the application and on the software/hardware
architecture.

However, it is clear that in many cases a deadline can be missed
without causing catastrophic consequences, but just with a simple
degradation of the performance of the system. Furthermore, as we will
see in Section~\ref{sec:control}, relaxing the hard
real-time constraint (and hence allowing some deadline misses) may
sometimes \emph{improve} the performance of a control system.

One may ask why real-time research literature is so much focused on
hard real-time systems. The main reason is \emph{separation of
  concerns}: by imposing hard real-time constraints, we can assess the
``correctness'' of the system by just analysing few,
application-independent parameters (like WCETs, minimum inter-arrival
times of events, etc.)  without the need to take into account the
final application requirements. 

Hence, 
by translating the application
requirements into hard real-time parameters and constraints (periods
and deadlines) we can easily derive general laws about scheduling.
Of course, something similar can be done using a soft real-time task
model in which we constraints the number or the extent of a deadline
miss.

\paragraph{Organisation of this paper}

The paper is organised as follows. In Section~\ref{sec:sys-model} we
present the model of a real-time task and the problem of large
variations in execution times. In Section~\ref{sec:soft} we briefly
present the Resource Reservation framework and the Constant Bandwidth
Server algorithm. We also discuss a model of soft real-time tasks and
the typical requirements. In Section~\ref{sec:soft-analysis} we give a
quick overview of the existing techniques and tools for soft real-time
scheduling and analysis. In Section~\ref{sec:control} we present a
technique for designing robust and efficient control systems based on
resource reservation techniques for coping with uncertainty in
execution times of control tasks. Finally, in Section~\ref{sec:concl}
we present our conclusions.

\section{System model}
\label{sec:sys-model}

\subsection{Hard real-time tasks}
\label{sec:hard-task-model}

In this paper we will focus on the classical model of periodic and
sporadic real-time tasks. However, many of the techniques presented
later are valid for more complex task models.

A real-time tasks $\tau_i$ is characterised by a tuple $\tau_i = (C_i,
D_i, T_i)$, where $C_i$ is the \emph{worst-case execution time}, $D_i$
is the \emph{relative deadline}, and $T_i$ is the \emph{minimum
  interarrival time}. A task produces a (finite or infinite) sequence
of jobs $J_{i,0}, J_{i,1}, \ldots$, and each job $J_{i,j}$ is
characterised by an absolute activation time $a_{i,j}$, a computation
time $c_{i,j}$, and an absolute deadline $d_{i,j}$. For a periodic
task $T_i$ represents the distance between two consecutive
activations, hence it must hold that $\forall j \geq 0, a_{i,j+1} =
a_{i,j} + T_i$, whereas for a \emph{sporadic task} $T_i$ represents
the \emph{minimum interarrival time}, so it must hold that $a_{i,j+1}
\geq a_{i,j} + T_i$. The absolute deadline is computed as $d_{i,j} =
a_{i,j} + D_i$. By the definition of WCET, it must hold that $\forall
j, c_{i,j} \leq C_i$.

A real-time tasks models a \emph{recurrent thread} in a real-time
operating system. Using the pthread library in Linux, the typical
structure of the code for a periodic task is represented in Listing
\ref{lst:periodic}.

\begin{lstlisting}[style=customc,gobble=4,caption={Structure of a periodic thread in Linux},label={lst:periodic}]
    struct per_data {
      int index;
      long period_us;
      long dline_us;
    };

    void *thread_code(void *arg) {
      struct per_data *ps = (struct per_data *) arg;

      // Initialization

      struct timespec next, dline, now;
      clock_gettime(CLOCK_REALTIME, &next);
      while (1) {

        // Job execution

        // Check deadline miss
        clock_gettime(CLOCK_REALTIME, &now);
        dline = next;
        timespec_add_us(&dline, ps->dline_us);
        if (timespec_cmp(&now, &dline) > 0) 
          pthread_exit(0);
       
        // Wait until next period
        timespec_add_us(&next, ps->period_us);
        clock_nanosleep(CLOCK_REALTIME, 
                        TIMER_ABSTIME, &next, NULL);
      }
      return NULL;
    }
\end{lstlisting}

The \lstinline!struct per_data! contains the period and the deadline
expressed in microseconds. After initialization the thread enters a
while loop where it executes the code of the jobs. At the end of the
execution, we check if the deadline has been missed, in which case we
abort the thread. Otherwise the thread suspends itself waiting for the
next periodic activation (at time \lstinline!next!).

In this example, we chose to abort the thread when a deadline is
missed. Please note that we detect the deadline miss \emph{after} the
job has already completed its execution, beyond its deadline. This
means that other lower priority threads may have been delayed by the
overrun of this thread, and this may cause future deadline misses of
those other threads. Therefore, the simple strategy shown in Listing
\ref{lst:periodic} may or may not be the appropriate. 

\subsection{Scheduling}
\label{sec:scheduling}

We consider an hardware platform system consisting $m$ identical
processors on which we execute the RTOS and the tasks. The task
execution is controlled by a \emph{scheduler} that decides at each
instant which tasks can be executed on the processors. Scheduling
algorithms can be classified into \emph{partitioned schedulers} and
\emph{global schedulers}. In the first case, tasks are statically
allocated to processors, and on each processor a scheduling algorithm
decided which of the tasks pertaining on that processor must be
executed. In global scheduling, every task can execute on any
processor. Typically, ready tasks are inserted into a single
\emph{ready queue} and the scheduling algorithm selects which task to
execute at each instant.


The most popular real-time scheduling algorithm is by far Fixed
Priority (FP), because it is implemented in every OS from nano-kernels
to large General Purpose OS like Windows and Linux.  In FP, tasks are
assigned fixed priorities and every job of that task has the same
priority.  The Earliest Deadline First (EDF) scheduler sorts tasks in
the ready queue according to their current absolute deadlines, and
executes the ones with the earliest deadline (hence the most
``urgent'').  EDF is less widespread but it is gaining popularity
especially for soft-real-time systems. An implementation of the EDF
with the Constant Bandwidth Server (see Section \ref{sec:soft}) is now
available in Linux starting from version
3.14~\cite{lelli2011efficient}.

A \emph{schedulability test} is an algorithm that, given a set of
tasks and a scheduling algorithm, tells us if during execution every
task will respect its deadlines. For hard real-time systems, it is of
paramount importance to run the test before deploying the software, to
guarantee that every deadline will always be met. A schedulability
test for soft real-time systems can give useful indications about the
typical soft-real-time constraints as the number of deadline misses,
the maximum tardiness (that is, the maximum extent of execution after
the deadline has expired), etc.

\subsection{Execution times}
\label{sec:exec-timevariation}

Executions times of tasks may vary. The variation depends on many
different factors: the algorithm implemented by the task, the input
data, the hardware architecture. In particular, the presence of
caches, the instruction pipeline, out-of-order execution of
instructions, etc. With the recent advances in processors and in
multicore systems, the impact of the architecture on the variation of
execution time has become larger and larger, to the point that the
worst-case execution time of a task may be orders of magnitude larger
than its average execution time, but it may happen very rarely. In
this case, a schedulability analysis based on WCETs leads to a very
low utilisation of resources.

Let see an example of what can happen when a task executes more than
expected. Consider a system with 3 tasks, $\tau_1=(1,4)$,
$\tau_2=(2,5)$, $\tau_3=(2,6)$, scheduled by EDF on a single processor
system. A simple utilization test for EDF consists on computing the
overall load of the system as $U = \sum_{i=1}^3 \frac{C_i}{T_i}$. If
all tasks have relative deadline equal to the period, and they are all
independent of each other, all tasks are schedulable if and only if $U
\leq 1$. In the example, $U=0,983$, so the system is
schedulable. However, due to a wrong estimations, it may happen that
the computation time of $\tau_1$ sometimes raises to $2$. In this
case, the deadline of \emph{any} instance could be missed.  

\begin{figure}
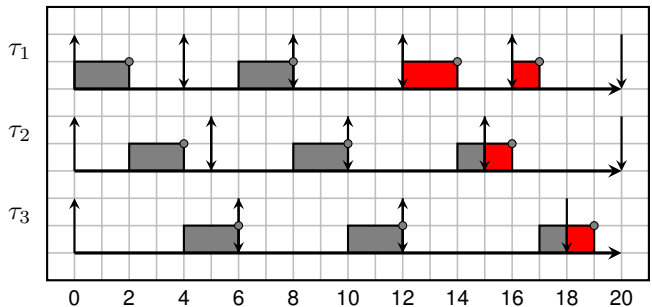

  \centering
  \begin{RTGrid}[width=8cm,numbersize=\footnotesize]{3}{20}
    \TaskNArrDead{1}{0}{4}{4}{5}
    \TaskNArrDead{2}{0}{5}{5}{4}
    \TaskNArrDead{3}{0}{6}{6}{3}
    \TaskExecDelta{1}{0}{2}
    \TaskEnd{1}{2}

    \TaskExecDelta{2}{2}{2}
    \TaskEnd{2}{4}
    
    \TaskExecDelta{3}{4}{2}
    \TaskEnd{3}{6}

    \TaskExecDelta{1}{6}{2}
    \TaskEnd{1}{8}

    \TaskExecDelta{2}{8}{2}
    \TaskEnd{2}{10}

    \TaskExecDelta{3}{10}{2}
    \TaskEnd{3}{12}

    \TaskExecDelta[color=red]{1}{12}{2}
    \TaskEnd{1}{14}

    \TaskExecDelta{2}{14}{1}
    \TaskExecDelta[color=red]{2}{15}{1}
    \TaskEnd{2}{16}

    \TaskExecDelta[color=red]{1}{16}{1}
    \TaskEnd{1}{17}

    \TaskExecDelta{3}{17}{1}
    \TaskExecDelta[color=red]{3}{18}{1}
    \TaskEnd{3}{19}

  \end{RTGrid}
  \caption{Example of overlaod for EDF.}
  \label{fig:edf-overload}
\end{figure}

In Figure~\ref{fig:edf-overload} we show the resulting EDF schedule
when the first 3 instances of $\tau_1$ execute for $2$ units, instead
of $1$. Each line represents the execution of one task. The up arrows
denote activation times, while downward arrows represents deadlines
(here deadlines are equal to periods, so all absolute deadlines
coincide with the next arrival time). The execution of every task is
represented by a grey rectangle if it is performed within the
deadline, and by a red rectangle if performed after the deadline. The
completion of an instance is denoted by a small black point in the
upper right corner of the rectangle.  As you can see, all three tasks
miss their deadlines.

To mitigate the problem, we could try to abort the third instance of
task $\tau_1$, but this does not guarantee that the other instances
will meet their deadlines, unfortunately: for example, the fourth
instance of the first task will still miss its deadline even when
aborting its third instance.

Using FP instead of EDF, only lower priority tasks will miss their
deadlines, therefore we achieve a sort of elementary temporal
isolation: high priority tasks are not influenced by lower priority
tasks. However, priority is not always related to criticality: for
example, to optimise resource utilisation the designer may choose to
set priorities in Rate Monotonic order (higher priority to tasks with
shorter periods) regardless of their criticality.

This problem has been addressed in many different ways. Recently, the
Mixed Criticality
methodology~\cite{vestal2007preemptive} has
been proposed: an an application consists of multiple criticality
levels, and different levels of assurance are provided to each
level. Typically, high criticality tasks are assigned an execution
budget, and when this budget is exceeded, the system goes into
\emph{high criticality mode} where all low criticality tasks are
discarded. This permits to guarantee the execution of higher
criticality tasks under all conditions, however it does not allow us
to control what happens to low criticality tasks.

Another approach is to isolate the temporal behaviour of each task by
using the Resource Reservation approach, which we describe in detail
in Section~\ref{sec:soft}.

\section{Resource reservations}
\label{sec:soft}

In the Resource Reservation framework~\cite{rajkumar1997resource},
every task is assigned to a \emph{scheduling server}, which is
characterised by a maximum \emph{execution budget} $Q$ and a period
$P$. The idea is that, if a task executes more than expected, it will
suffer but it will not negatively influence the execution of the other
tasks. One popular resource reservation algorithm is the Constant
Bandwidth Server~\cite{abeni1998integrating} which integrates with the
EDF scheduler. Basically, the algorithm rules say that the execution
budget is recharged at the beginning of each period, and it is
decreased when the task executes. If the execution budget reaches 0
before the task has completed its execution, the task deadline is
postponed by $P$ units of time, so its dynamic priority is lowered and
the budget is recharged to $Q$. In another version (called \emph{hard
  reservation}), the task whose budget is exhausted is suspended until
the end of the period when its budget is finally recharged to $Q$.
For more details on the algorithm, please refer to the original
paper. Resource reservation provide the important \emph{temporal
  isolation} property: the ability of a task to meet its deadlines
does not depend on the behaviour of the other tasks, but only on its
execution time profile and on the budget and period assigned to it. In
addition, the CBS provides the \emph{hard schedulability} property: is
we assigned a budget larger than the WCET of the task, and a
reservation period no larger than the task's period, the task is
guaranteed to meet all its deadlines.

The algorithm has been recently implemented in Linux, and starting
from version 3.14, it is bundled in the official Linux distribution. A
new scheduler is available to developers, (called
\texttt{SCHED\_DEADLINE}) and the superuser can use it to assign
budget and periods to any Linux thread.

\begin{figure}
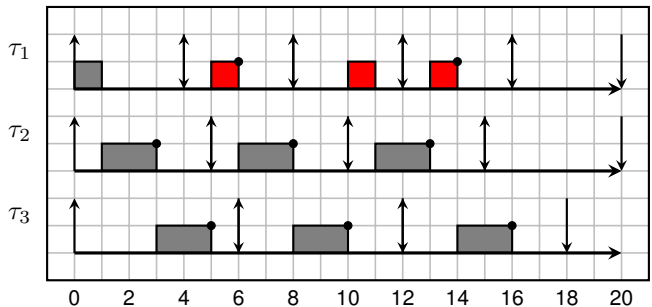

  \centering
  \begin{RTGrid}[width=8cm,numbersize=\footnotesize]{3}{20}
    \TaskNArrDead{1}{0}{4}{4}{5}
    \TaskNArrDead{2}{0}{5}{5}{4}
    \TaskNArrDead{3}{0}{6}{6}{3}

    \TaskExecDelta{1}{0}{1}

    \TaskExecDelta[end=1]{2}{1}{2}
    
    \TaskExecDelta[end=1]{3}{3}{2}

    \TaskExecDelta[end=1,color=red]{1}{5}{1}

    \TaskExecDelta[end=1]{2}{6}{2}

    \TaskExecDelta[end=1]{3}{8}{2}

    \TaskExecDelta[color=red]{1}{10}{1}

    \TaskExecDelta[end=1]{2}{11}{2}

    \TaskExecDelta[end=1,color=red]{1}{13}{1}

    \TaskExecDelta[end=1]{3}{14}{2}
  \end{RTGrid}
  \caption{Same example scheduled by the CBS.}
  \label{fig:cbs-overload}
\end{figure}

The same example of Figure \ref{fig:edf-overload} is scheduled in
Figure \ref{fig:cbs-overload} using the CBS. Task $\tau_1$ is assigned
a server $S_1 = (Q=1,P=4)$, task $\tau_2$ is assigned $S_2 =
(Q=2,P=5)$, and task $\tau_3$ is assigned $S_3=(Q=2,P=6)$. Task
$\tau_1$ is thus constrained to execute only $1$ unit of execution
time every 4, and it will miss many deadlines, whereas the other two
tasks are not influenced.

\begin{figure}
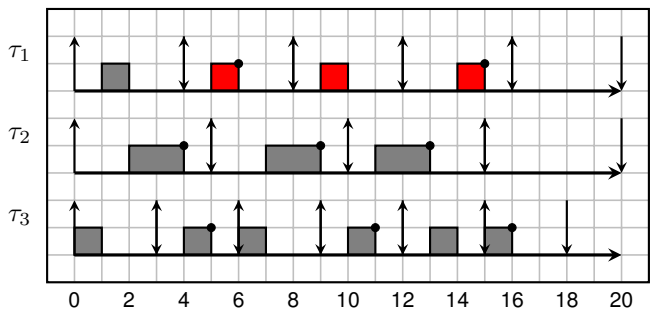

  \centering
  \begin{RTGrid}[width=8cm,numbersize=\footnotesize]{3}{20}
    \TaskNArrDead{1}{0}{4}{4}{5} 
    \TaskNArrDead{2}{0}{5}{5}{4}
    \TaskNArrDead{3}{0}{3}{3}{6}

    \TaskExecDelta{3}{0}{1}

    \TaskExecDelta{1}{1}{1}

    \TaskExecDelta[end=1]{2}{2}{2}
    
    \TaskExecDelta[end=1]{3}{4}{1}

    \TaskExecDelta[end=1,color=red]{1}{5}{1}

    \TaskExecDelta{3}{6}{1}

    \TaskExecDelta[end=1]{2}{7}{2}

    \TaskExecDelta[end=1]{3}{10}{1}

    \TaskExecDelta[color=red]{1}{9}{1}

    \TaskExecDelta[end=1]{2}{11}{2}

    \TaskExecDelta{3}{13}{1}

    \TaskExecDelta[end=1,color=red]{1}{14}{1}

    \TaskExecDelta[end=1]{3}{15}{1}
  \end{RTGrid}
  \caption{Task $\tau_3$ is assigned a server with half its period.}
  \label{fig:cbs-half-period}
\end{figure}

It is important to underline that the period of the server must not be
necessarily equal to the period of the task. In fact, sometimes it may
be appropriate to set $P=\frac{T}{k}$, with $k$ positive small
integer. If we set $k$ to a large integer number, the execution of the
task will be \emph{spread} over its period, and the larger is $k$ the
more the execution resemble a \emph{fluid execution}. However a large
$k$ implies a large number of context switches and hence a much larger
overhead of the scheduler. Using a small $k$ is useful for controlling
precisely the timing of the task output, as will be explained in
Section~\ref{sec:control}. In Figure~\ref{fig:cbs-half-period} we show
the same example as above, where task $\tau_3$ is assigned a server
with $S_3=(Q=1, P=3)$.

\subsection{Soft real-time tasks}
\label{sec:soft-model}

The model for soft real-time tasks is very similar to the model for
hard real-time tasks. There are two main differences: 1) what to do
when a deadline is missed and 2) the notion of \emph{correctness},
that is what are the requirements of a soft real-time tasks.

We start by defining the \emph{tardiness} of a task as the maximum
delay in completing an instance after its deadline. If we denote by
$R_{i,j}$ the response time of the j-th instance of task $\tau_i$, its
tardiness is defined as $\max_j(0, R_{i,j} - D_i)$. Of course, hard
real-time tasks must have tardiness equal to 0.

There are many ways to handle deadline misses and overruns of soft
real-time tasks, here we summarise the major ones.

\begin{itemize}
\item \textbf{Deadline miss detection}: the program detects that a
  deadline has been missed, and takes the appropriate action. We can
  detect the occurrence of a deadline miss at the end of the job (as
  in Listing~\ref{lst:periodic}) or at the time of the deadline miss:
  in any case, we are already executing after the deadline, so we have
  to take some recovery action to remove the overload
  situation. Notice that, if there is no temporal isolation, other
  tasks may miss their deadlines in cascade. In case of resource
  reservations, instead, only the failing task will be concerned.

\item \textbf{Budget overrun detection}: in this case, the system
  informs us when the task is exceeding its execution budget. We can
  then suspend the task and let it continue later on (as in the
  resource reservation framework); or abort the task instance, so that
  it can start clean again in next period.
\end{itemize}

It is important to underline that the second technique (budget overrun
detection) is widely used also in safety critical hard real-time
systems for fault-tolerant reasons.

What to do when a deadline miss or a budget overrun is detected?
again, there are many possibilities:
\begin{itemize}
\item We can \textbf{abort} the executing instance and use some standard value
  of the task output. This can be done to alleviate the overload of
  the system and while providing an output to the actuators (or to the
  following tasks in the chain). However, we must be careful in
  aborting an executing thread: the thread must release all mutually
  exclusive semaphores before aborting, and it must leave all data
  structures in a consistent state. This is not always easy to
  achieve.

\item We can \textbf{continue} the task execution and finish
  late. Sometimes a late result is better than no result, especially
  if we can prove that the overall tardiness is limited. Also, it is
  possible that the overload will decrease by itself is the system is
  not permanently overloaded.
  
\item As an alternative, we can decide to \textbf{skip} future
  instances, so to decrease the overload and keep the maximum
  tardiness in check. Also in this case, we still need to provide
  standard values for the outputs of the skipped instances.
\end{itemize}

Which output should be provided for aborted or skipped instances? A
typical strategy is to hold the previous value of the output, both in
control applications and in multimedia applications. In the first
case, we need to model this fact in the design of the controller. In
the second case, for video decoders the image on the screen may freeze
for an instant, and hopefully return to normal operation in the
following frames; for audio applications this can result in a noise,
so it is sometimes better to send a forecast signal or an empty
signal.

In Listing~\ref{lst:skip} we show an example of how to combine these
techniques by using \texttt{SCHED\_DEADLINE} for budget control, and
by skipping all late instances if a deadline has been missed (lines
18-20).

\begin{lstlisting}[style=customc,gobble=4,caption={Structure of a periodic thread in Linux},label={lst:skip}]
    void *thread_code(void *arg) {
      struct per_data *ps = (struct per_data *) arg;

      struct timespec next, now;
      clock_gettime(CLOCK_REALTIME, &next);
      while (1) {
       // Wait until next period
       timespec_add_us(&next, ps->period_us);
       clock_nanosleep(CLOCK_REALTIME, TIMER_ABSTIME, 
                       &next, NULL);

        // Job execution

        // Check deadline miss
        clock_gettime(CLOCK_REALTIME, &now);
        while (timespec_cmp(&now, &next) > 0) {
          // Skip late instances
          timespec_add_us(&next, ps->dline_us);
        }
      }
      return NULL;
    }
\end{lstlisting}

\subsection{Soft real-time requirements}
\label{sec:soft-requirements}

Even if deadlines can sometimes be missed, we need to do an off-line
analysis of the system to estimate the number and frequency of missed
deadlines. We also need on-line techniques to keep these misses in
check. 
Typically, we have three types of requirements:

\begin{itemize}
\item Bound on the \textbf{number of deadline misses in an interval of
    time}. We express this constraint as $\binom{m}{n}$, meaning that
  we can miss at most $m$ deadline over $n$ instances. This is
  sometimes used in the design of control systems to ensure stability
  of the system. For example, we may require that, over 10 instances, no
  more than 2 will miss their deadlines as $\binom{2}{10}$. We can
  also impose several of these constraints: for example, we want at
  most 2 deadline misses but never consecutively. In this case, we
  express the constraint as $\binom{2}{10} \wedge \binom{1}{2}$.

\item Bound on the \textbf{tardiness} of a task. Task's instance can
  complete after their deadlines, but the delay must be bounded. Of
  course, this makes sense only if we decide to not abort the instance
  in case of a deadline miss.

\item \textbf{Probabilistic bounds}: the same requirements can be
  expressed in terms of probability. For example, we look for upper
  bounds on the probability of having a deadline miss. Clearly,
  probabilistic analysis needs a probabilistic characterisation of the
  execution time of a task.
\end{itemize}

\section{Algorithms for soft real-time tasks}
\label{sec:soft-analysis}

The resource reservation framework provide the \emph{temporal
  isolation} property which is useful to analyse each soft real-time
task in isolation. This simplifies the analysis and the algorithms for
dealing with variations in execution times. Many analysis algorithms
and scheduling methods have been proposed in the literature for
dealing with soft real-time tasks. In the following sections we review
some of the most relevant papers on analysis of soft real-time tasks
in the context of resource reservations.

\subsection{Probabilistic analysis}
\label{sec:probability}

If we want to analyse a soft real-time task with large variations in
execution times, one approach is to reason in terms of probability. An
initial proposal in this regard was made by Abeni and
Buttazzo~\cite{abeni1999qos}. They proposed a methodology for
computing the probability of finishing time of a soft real-time task
served by a CBS with given budget and period, starting from the
probability distribution of the execution times. Recent works by
Palopoli et al. \cite{ECRTS12,TPD13,abeni2012efficient} have refined
the initial analysis providing analytical bounds on the deadline miss
probability.

\subsection{Reclaiming}
\label{sec:reclaiming}

Typically, a soft real-time task is assigned a server with budget not
inferior to its average execution time. This means that, when the task
needs to execute more than its assigned budget, it will probably miss
its deadline. However, when the task needs to execute for less than
its budget, the remaining part is discarded. Therefore, it is
interesting to see if it is possible to redistribute the unused budget
to the other more needing tasks, so to reduce the overall probability
of deadline miss.

Many algorithms have been proposed in the literature for implementing
budget reclaiming mechanisms. In this paper we will limit ourselves to
the methods that extend the CBS server.

The CASH (Capacity Sharing) algorithm \cite{caccamo2000capacity}
recuperates the unused budgets in a queue. The queue contains pairs
$(q_r, d_r)$, where $q_r$ is the remaining budget, and $d_r$ is the
deadline of the corresponding server. The unused budget can be donate
to a server with scheduling deadline not earlier to $d_r$. This method
works well for periodic tasks but it is difficult to extend to
sporadic tasks. The CASH algorithm has later been extended to
multiprocessor systems \cite{pellizzoni2008m}.

The GRUB (Greedy Reclamation of Unused Bandwidth) algorithm
\cite{conf/ecrts/LipariB00} uses an analogy with a fluid scheduler to
keep track of the \emph{active bandwidth} of the system $U_{act}$. The
remaining bandwidth $1-U_{act}$ is donated to the server currently
executing. It works for any kind of task (periodic, sporadic and even
aperiodic). A variant called ShRUB (Shared Reclamation of Unused
Bandwidth) \cite{DBLP:conf/estimedia/PalopoliACLB08} distributes the
spare bandwidth across all active servers. GRUB has been extended to
power-aware scheduling \cite{journals/tc/ScordinoL06} since reclaiming
and power-aware scheduling are two faces of the same problem. The GRUB
algorithm and its variants have been designed for single processor
systems. 

Other relevant works in this area include
\cite{lin2005improving,nogueira2007capacity,marzario2004iris}.

\subsection{Adaptive reservations}
\label{sec:adaptive}

When using resource reservations, one difficult problem to solve is
how to assign the server's budget. A too small budget leads to many
deadline misses, whereas a too large budget wastes precious
resources. In addition, tasks may exhibit structural variations in
execution times. For example, a task may run one of two different
algorithms, depending on the \emph{operational mode} of the system,
and these two algorithms may exhibit completely different execution
time probability distributions.

Therefore, one interesting approach is to \emph{adapt} the budget to
the needs of the task at on-line. This can be done by using a feedback
control scheme: while the task executes, the \emph{scheduling error}
(that is the tardiness, or the number of deadline misses) is measured,
and a feedback control algorithm adjusts the budget in order to reduce
the scheduling error to zero.

\emph{Adaptive reservations} have been proposed in
\cite{abeni2002analysis} and later in
\cite{abeni2005qos,cucinotta2011robust,DBLP:journals/spe/PalopoliCML09}
as a mean to support feedback based adaptive budget control. In
particular, implementations have been proposed for the Linux OS in
\cite{cucinotta2004adaptive,DBLP:journals/spe/PalopoliCML09} for
dealing with multimedia applications as video players. If a model of
the application is known (for example, a model of the MPEG decoder),
it is possible to give guarantees about the stability of the
controller and on the QoS. A complete QoS architecture has been
proposed for single processors and for distributed systems
\cite{DBLP:journals/tii/CucinottaPAFL10} and for industrial automation
applications \cite{DBLP:journals/tii/CucinottaMALMCR09}.

For multimedia applications that have not been designed according to
the soft real-time paradigm (but they still exhibit real-time
requirements), it is possible to automatically detect the best
``period'' of the reservation server, using simple filters inside the
OS. This idea has been implemented in a mechanism for the Linux kernel
\cite{cucinotta2012adaptive}. 

\subsection{Shared resources}
\label{sec:extensions}

If a task can block due to a locked semaphore on a shared resource,
its remaining budget is not valid anymore. Moreover, if a task that
holds as shared resource is suspended because its budget has been
exhausted, blocked may suffer a form of priority inversion that can
seriously compromise schedulability.  Therefore, it is necessary to
modify the reservation algorithm and the synchronisation protocol to
take into account these interactions.

There are three main approaches to do this. 

\begin{itemize}
\item In the first, we maintain the reservation algorithm unchanged
  but we check if there is enough budget left before entering a
  critical section: if the length of the critical section is larger
  than the remaining budget, the task is suspended. In this way we
  avoid the second problem. Blocking times are then calculated using
  the synchronisation protocol (e.g. Stack Resource Policy or Priority
  Inheritance) and taken into account into the schedulability
  test. This is the approach followed by the BROE
  \cite{bertogna2009resource} algorithm.
\item Another approach is to avoid suspending the task if is runs
  inside a critical section. The overrun are accounted for in the
  analysis together with the blocking times. This is the approach
  followed by the SIRAP algorithm \cite{behnam2007sirap}.

\item The third approach is to use \emph{bandwidth inheritance}: if a
  task blocks on a critical section, the holding task can inherit the
  budget of the blocked task, and use the budget with the shorter
  deadline. This is the approach of the BWI (Bandwidth Inheritance)
  Protocol \cite{lipari2004task} and its multiprocessor extension M-BWI
  \cite{Faggioli2012}. This last protocol does not require any
  specific knowledge of the length of the critical sections (such a
  knowledge is needed only for analysis) and can still guarantee
  temporal isolation between non interacting tasks.
\end{itemize}

In our opinion, the latter is the most adequate for soft real-time
systems, as the isolation property does not depend on the correct
estimation of the length of the critical sections and of the blocking
time. Also, when adding a new task/reservation in the system, it is
not necessary to recalculate the resource holding times. We are
currently working on a comparison of the different approaches in terms
of resource utilisation and implementation details.

\section{Control systems: hard or soft? }
\label{sec:control}

\subsection{Generalities on control systems}
\label{sec:gener-contr-syst}

A feedback control system is given by the interconnection of a plant
(a ``physical'' system to be controlled) and of a digital controller.

The plant is typically described by a differential equation:
\begin{align*}
\dot{x}(t) &= f(x(t), u(t), t)\\
y(t) &= g(x(t), u(t)),
\end{align*}
$t \in \R$ represents time, $x(t) \in \R^n$ is a vector of state variables, 
$y(t) \in \R^m$ is a vector of output variables and $u(t) \in \R^p$ is a set
of input variables whereby the system evolution can be controlled.
For the purposes of this paper, we will restrict our attention to linear
and time invariant plants, for which the differential equation above can be
specialised as 
\begin{align}
\dot{x}(t) &= A_c x(t) + B_c u(t) \nonumber\\
y(t) &= C_c x(t) + D_c u(t), \label{eq:LTI-system}
\end{align}
where $A$, $B$, $C$ and $D$ are matrices of suitable size.  Typical
tasks of a control system are to steer the system state to a desired
equilibrium $x_{\text{eq}}$ (point stabilisation) or along a specified
trajectory $y_{ref}(t)$ (tracking).  Instrumental to this goal is the
solution of a particular sub-problem, which is the stabilisation of
the origin of the state space (i.e., driving the system state $x(t)$
to $0$). This problem is called \emph{stabilisation}, and in this paper
will receive a special attention.

The stabilisation problem amounts to finding a command value $u(t)$
such that the resulting \emph{closed loop} system is asymptotically
stable meaning that: 1) for all $\epsilon$ we can find a $\delta$ such
that if the initial state $x(0)$ is within a distance $\delta$ of the
origin, then $x(t)$ will always be within a distance $\epsilon$ for
all $t$, 2) $\lim_{t \rightarrow \infty} x(t) = 0$.
In plain words, if the system evolution starts close to the origin, it will 
always remain close to the origin, and it will eventually converge 
to the origin itself.

Different design methods can be found that achieve asymptotic stability and
that possibly fulfil additional requirement~\cite{goodwin2001control}.
Generally speaking, such methods produce a controller, which is itself a dynamic
system that receives as input the output $y(t)$ of the plant and produces as
output the input $u(t)$ for the plant
\begin{align}
\dot{z}(t) &= E_c z(t) + F_c y(t) \nonumber\\
u(t) &= G_c z(t) \label{eq:LTI-controller}
\end{align}
The vector $z$ is compounded by the state variables of the controller
and it can have a different size according to the type of controller
chosen. For instance, for LQG controller $z$ has the same size of the
state vector $x$ of the plant. The LQG controller has the important
property: if the system is acted on by a noise term (i.e., in
adddition to $u$ we have another uncontrollable input $w$ that
fluctuates stochastically), the controller minimises a cost function
like $\int_0^\infty E\{ x^T Q x + u^t R u dt \}$, which accounts for
the evolution of the state variables and for the control action.

The digital implementation of a controller like this amounts to
developing a program that cyclically: 1) takes a sample of the output
$y$ of the plant, 2) updates the state of the controller producing an
approximation of the solution of the differential
Equation~\ref{eq:LTI-controller}, 3) computes a command value $u$ that
is applied to the plant and held constant until the next sample is
ready.  The name of the game is to find a technological scheme such
that the behaviour of the ``ideal'' control in
Equation~\ref{eq:LTI-controller} is closely approximated and its
relevant properties (first and foremost stability) preserved.  The
classic results on digital control~\cite{franklin1998digital} show how
to achieve the result under the assumptions that the loop is executed
with a fixed periodicity $T$ and that computation time can be
neglected. In practical terms, this is done by introducing a
discrete-time model that describes the evolution of 
plant and controller across a sampling period $[kT,\,(k+1)T]$, under
the typical assumption that control values are held constant
throughout:
\begin{align}
x ((k+1)T) &= A x(kt) + B u(kt) \nonumber\\
y(kT) &= C x(kT) + D u(k t) \nonumber \\
\nonumber \\
z((k+1)T) &= E z(kT) + F y(kT) \nonumber \\
u(kT) &= G_c z(kT) \label{eq:LTI-system2}        
\end{align}

Computation delays do not pose serious issues as long as they are
fixed. In this case the designer can easily account for the delays by
extending the model of the plant with additional state variables and
compensate for the dynamics of these additional variables in her/his
controller.

\subsection{Hard Real--Time implementation of controllers on multi-programmed systems}
\label{sec:hard-real-time}

Delay compensation cannot be done easily if the computation time of
the task has a wide variability. Things become even trickier when the
processor is shared with other tasks. In this scenario, the task can
suffer from scheduling interference introduced by preempting tasks
which adds to the variability of the introduced delays. A pragmatic
way to solve this problem is by using the so-called time-triggered
model of computation~\cite{kopetz2003time}. The idea is that sampling
and actuations are \emph{forced} to take place at specific points in
time, regardless of when each job of the task starts or ends.  For
instance, suppose that samples are collected periodically with period
$T$. \emph{Exactly} at time $kT$ with $k \in \N$, a new sample is
collected \footnote{This can be done with an appropriate hardware
  which ensures a small sampling jitter.} and a new job $J_k$ is
activated. The job is assigned a relative deadline $D$ to complete
its computation.  When the job finishes, the result is retained in an
intermediate buffer and is delivered to the actuators only at time $kT
+ D$.  The evident advantage of this idea is that jobs are allowed to
have a variable response time \emph{but} the sensor-to-actuator delay
is always $D$. The only requirement is that the deadline $D$ is
enforced for each and every job and this can be done using the
standard methods of hard real-time scheduling theory.

This idea is therefore based on a clear separation of concerns between
the control designer (who compensates for fixed delays) and the system
designer (who uses real-time scheduling to secure the timely execution
of the jobs).  An important problem with this ``clean'' approach comes
when the execution time of the control task changes by a significnt
amount on a job-by-job basis.  This is the exactly the case for a new
generation of control application that rely on data collected from
visual sensors, LIDARS and RADARS. The processing time required to
extract the meaningful information is heavily dependent on the
environment (e.g., extracting relevant features from a clean image is
obviously less demanding than from an images cluttered with all sort
of artefacts). For applications like these, guaranteeing every single
deadline under all possible conditions understandably requires a
significant over-provisioning of system resources. On the other hand,
several studies~\cite{Cer04,Pal00} reveal that even in control systems
a ``controlled'' fluctuation in the delays introduced by the feedback
loop can be tolerated without evident degradations in control
performance.  What is needed to translate this simple observation into
a credible alternative to hard real-time design is a rigorous
methodology with certifiable results.

\subsection{Making the case for soft real--time}
\label{sec:making-case-soft}

Any methodology for ``certifiable'' design of real-time embedded
controllers rests on three pillars:
\begin{enumerate}
\item A clear understanding of the system level properties that need
  to be guaranteed,
 
\item A model for the temporal behaviour of the computing platform
  that can be treated in control design,
 
\item A platform design based on a model of computation and a
  scheduling solution that allows to control the temporal behaviour of
  the tasks for an assigned choice of design parameters.
\end{enumerate}
For classic ``hard real-time'' methods the system level property that
needs to be preserved is usually related to system asymptotic
stability, the temporal behaviour of the task is captured by the
$(T,D)$ pair, where $T$ is the sampling time and $D$ is the sensor/to
actuator delay and the platform design is based on the combination of
time-triggered model of computation and fixed priority scheduling.
Any alternative paradigm has to propose a credible alternative to each
of these three elements.

Generally speaking, if we give up the strict hard real-time execution
guarantees, we are implicitly injecting a stochastic component related
to the behaviour of the controller.  The property required to the
closed loop system will have a stochastic flavour in its turn.  There
are different possibilities.  The first one is \emph{almost sure
  stability}, which essentially means that all the realisations but a
set of null measure will converge to $0$: $\text{Prob\{}\lim_{t
  \rightarrow \infty} |x(t)| = 0 \text{\}} = 1$.  A different
possibility is the so-called \emph{second moment stability}, meaning
that the variance of the state will converge to $0$: $\lim_{t
  \rightarrow \infty} E\left\{x^T(t) Q x(t) \right\} = 0$. The
intuition is that the state will be stocastically contained in a ball
of shrinking size. We will adopt this second notion in the discussion
below.

As for the platform design, the use of the CBS scheduler has the
remarkable property that a choice of the scheduling parameters is
immediately translated into the probability distribution of having
different delays in the computation~\cite{TPD13,ECRTS12}, which is
utmost useful for control design. For the model of computation,
different choices are possible with different results, as discussed
next.

\subsubsection{A first possibility: job dropout}
\label{sec:first-poss-job}

The simplest possible choice for a model of computation is to adopt a
standard Time Triggered model that acquires sensor input exactly at
the start of a sampling period (e.g., $kT$) and releases the output
exactly at the end (e.g., $(k+1)T$) with the addition that if the job
executes beyond $kT$ it is outright cancelled and the past input is
held constant throughout the next job (from $(k+1)T$ to $(k+2)T$). In
this case the evolution of the total state
$\hat{x}=\begin{smallmatrix} x \\ z \end{smallmatrix}$, comprised of
the state of the plant ($x$) and of the controller $z$, is given by:
{\small
\begin{align}
\hat{x}((k+1)T) &= \begin{cases} A_c \hat{x}(kT) & \text{if the
    $k^{th}$ job finished within $(k+1)T$ }\\
A_o \hat{x}(kT)  & \text{if it does not }\\ 
\end{cases} \label{eq:switching}
\end{align}} 
for an appropriate expression of the $A_c$ and $A_o$
matrices (see~\cite{fontantelli2013optimal}).

With the simple Model of Computation proposed here, there is not any
``carry-on'' execution across to ajacent jobs. Therefore the
stochastic process ruling the transition between the two modes
(deadline respected and deadline missed) is memoryless.  Thanks to the
adoption of the Resource Reservation algorithm, the probability $\mu$
of dropping a job (with the system evolving according to the ``open
loop'' matrix $A_o$) can easily be found as
$\text{Prob}\left\{\left\lceil \frac{c_j}{Q} \right\rceil R >
  T\right\rceil$ where $Q$ is the reservation budget, $R$ is the
reservation period, $T$ is the task period and $c_j$ is the
(stochastic) computation time of the $j^{th}$ job.  For a given
probability distribution for $c_j$, it is straightforward to find this
probability as a function of $Q$ and $R$.  Further, being the
transition between the two modes ruled by a memoryless process, the
second moment stability is expressed by the relatively simple
requirement that the matrix
\begin{align}
\tilde{A}= (1-\mu) A_c \otimes A_c + \mu A_o \otimes A_o
\label{eq:stability-matrix}
\end{align}
has all eigenvalues with modulus smaller than $1$, where $\otimes$ denotes the Kronecker
product.

In summary, by using CBS with a simple model of computation where
delayed jobs are simply dropped, it is easy to establish a link between
the scheduling parameters and the probability $\mu$ of dropping a job,
and also to find if the resulting system is second moment stable.  The
price to pay is that the outright cancellation of a delayed job can be
a very crude choice and system stability can require unnecessarily
high values of the bandwidth.

\subsubsection{A Better Alternative}
\label{sec:better-alternative}

The problem with the approach above is that the system could possibly
benefit more from a fresher control output (although generated with a
little delay) than from an abrupt cancellation.  On the other hand,
the simplification of constraining the possible delays to one (or at
least) a few possible values is too important to be easily dismissed.

\begin{figure}
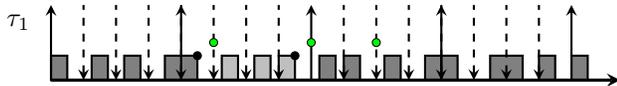


  \begin{RTGrid}[nogrid=1,nonumbers=1,width=8cm,height=2cm,exeheight=.65]{1}{35}
    \TaskNArrival{1}{0}{8}{5}
    
    \TaskExecDelta{1}{0}{1}
    \TaskExecDelta{1}{2.5}{1}
    \TaskExecDelta{1}{4.5}{1}
    \TaskExecDelta[end=1]{1}{7}{2}

    \TaskDeadline[style=dashed]{1}{2}
    \TaskDeadline[style=dashed]{1}{4}
    \TaskDeadline[style=dashed]{1}{6}
    \TaskDeadline[style=dashed]{1}{8}

    \TaskDeadline[style=dashed]{1}{10}
    \TaskDeadline[style=dashed]{1}{12}
    \TaskDeadline[style=dashed]{1}{14}

    \TaskEnd[color=green]{1}{10}

    \TaskExecDelta[color=lightgray]{1}{10.5}{1}
    \TaskExecDelta[color=lightgray]{1}{12.5}{1}
    \TaskExecDelta[end=1,color=lightgray]{1}{14}{1}

    \TaskEnd[color=green]{1}{16}

    \TaskDeadline[style=dashed]{1}{18}
    \TaskDeadline[style=dashed]{1}{20}
    \TaskDeadline[style=dashed]{1}{22}

    \TaskDeadline[style=dashed]{1}{24}
    
    \TaskExecDelta{1}{16.5}{1}
    \TaskExecDelta{1}{18}{1}
    \TaskExecDelta{1}{20.5}{1}
    \TaskExecDelta{1}{23}{1}

    \TaskDeadline[style=dashed]{1}{26}
    \TaskDeadline[style=dashed]{1}{28}
    \TaskDeadline[style=dashed]{1}{30}

    \TaskExecDelta{1}{24}{1}
    \TaskExecDelta{1}{27}{2}
    \TaskExecDelta{1}{30}{1}
    \TaskExecDelta{1}{32}{1}

    \TaskEnd[color=green]{1}{20}

  \end{RTGrid}

\caption{Example schedule for soft real--time control task}
\label{fig:soft-rt-sched}
\end{figure}

One possibility before us is to use the combination of a real-time
tasking model (where job activations are buffered until the previous
jobs finish), with a time-triggered model of computation where
the control input is collected with a fixed periodicity (release
points) and the
output is released only upon specified instants, which is convenient
to set equal to the end of the reservation periods. 
An example is shown in Figure~\ref{fig:soft-rt-sched}.
The dashed lines denote the acceptable release points for the output,
and the green circle represents the moment a new fresh output is
taken.  In this example, Job 1 is late. On the expiration of its
period, the input for Job 2 is collected, but Job 1 keeps going. When
Job 1 finishes its output is released upon the next possible point
(green circle) and Job 2 can start.  As regards Job 3, we can see that
the job is late beyond an acceptable threshold and is cancelled.

As discussed in out previous work~\cite{FontanelliGP13}, it is
possible to model the evolution of the delays as a Markov Chain, whose
particular structure simplifies the computation of the steady state
distributions~\cite{TPD13,ECRTS12}.  The resulting closed loop system
is a so called Makrov Jump Linear System (MJLP), and its second moment
stability can be analysed with (rather complex) mathematical tools.

To make a long story short, allowing for delays makes loosen the
unnecessary restrictions of the Model of Computation where late jobs
are cancelled. This can bring to important savings of
bandwidth. However, the ensemble task/scheduler becomes a system with
memory (a Markov Chain), which is certainly more difficult to analyse.

\subsubsection{An even better alternative}
\label{sec:cstream}
\begin{figure}
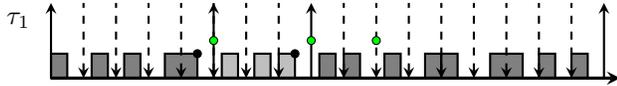


  \begin{RTGrid}[nogrid=1,nonumbers=1,width=8cm,height=2cm,exeheight=.65]{1}{35}
    \TaskArrival{1}{0}
    \TaskArrival{1}{10}
    \TaskArrival{1}{16}
    \TaskArrival{1}{34}

    \TaskExecDelta{1}{0}{1}
    \TaskExecDelta{1}{2.5}{1}
    \TaskExecDelta{1}{4.5}{1}
    \TaskExecDelta[end=1]{1}{7}{2}

    \TaskDeadline[style=dashed]{1}{2}
    \TaskDeadline[style=dashed]{1}{4}
    \TaskDeadline[style=dashed]{1}{6}
    \TaskDeadline[style=dashed]{1}{8}

    \TaskDeadline[style=dashed]{1}{10}
    \TaskDeadline[style=dashed]{1}{12}
    \TaskDeadline[style=dashed]{1}{14}

    \TaskEnd[color=green]{1}{10}

    \TaskExecDelta[color=lightgray]{1}{10.5}{1}
    \TaskExecDelta[color=lightgray]{1}{12.5}{1}
    \TaskExecDelta[end=1,color=lightgray]{1}{14}{1}

    \TaskEnd[color=green]{1}{16}

    \TaskDeadline[style=dashed]{1}{18}
    \TaskDeadline[style=dashed]{1}{20}
    \TaskDeadline[style=dashed]{1}{22}
    \TaskDeadline[style=dashed]{1}{24}
        
    \TaskExecDelta{1}{16.5}{1}
    \TaskExecDelta{1}{18}{1}
    \TaskExecDelta{1}{20.5}{1}
    \TaskExecDelta{1}{23}{1}

    \TaskDeadline[style=dashed]{1}{26}
    \TaskDeadline[style=dashed]{1}{28}
    \TaskDeadline[style=dashed]{1}{30}
    \TaskDeadline[style=dashed]{1}{32}

    \TaskExecDelta{1}{24}{1}
    \TaskExecDelta{1}{27}{2}
    \TaskExecDelta{1}{30}{1}
    \TaskExecDelta{1}{32}{1}

    \TaskEnd[color=green]{1}{20}

  \end{RTGrid}

\caption{Example schedule the Continuous Stream model}
\label{fig:cstream}
\end{figure}

As surprising as it can seem, we can allow for delays and yet have a
model of computation as easy to analyse as time-triggered with job
dropout. To do so we have to completely give up the idea of periodic
sampling and adopt an event-triggered sampling instead.  However,
contrary to other event-triggered alternatives proposed in the
literature the event we look does not emanate from the system we
control but from the controller itself: a new sample is acquired and a
new job is released only when the previous job terminates (to be
precise on the earliest release point available after the job
finishes. This model is called Continuous Stream and is described
in~\cite{FontanelliPA13}. An example schedule is shown in
Figure~\ref{fig:cstream}
As is it is possible to see, no new activation is triggered for Job 2
nor any sample is collected before Job 1 finishes. The same applies
for Job 3, which is cancelled because its execution exceeds a maximum
delay threshold.

The fact that a task starts only after the previous one finishes
cancels any carry on execution across two adjacent jobs and the
ensemble scheduler/task becomes memoryless. In other words, we never
have a job in the wait queue of a task that is still in execution.
The control theoretical consequence of this fact is that we can set up
the stability problem by requiring that a matrix $\tilde{A}$, as defined
in Equation~\ref{eq:stability-matrix}, has all eigenvalues in the unit
circle.
\begin{figure}
\includegraphics[width=0.9\columnwidth]{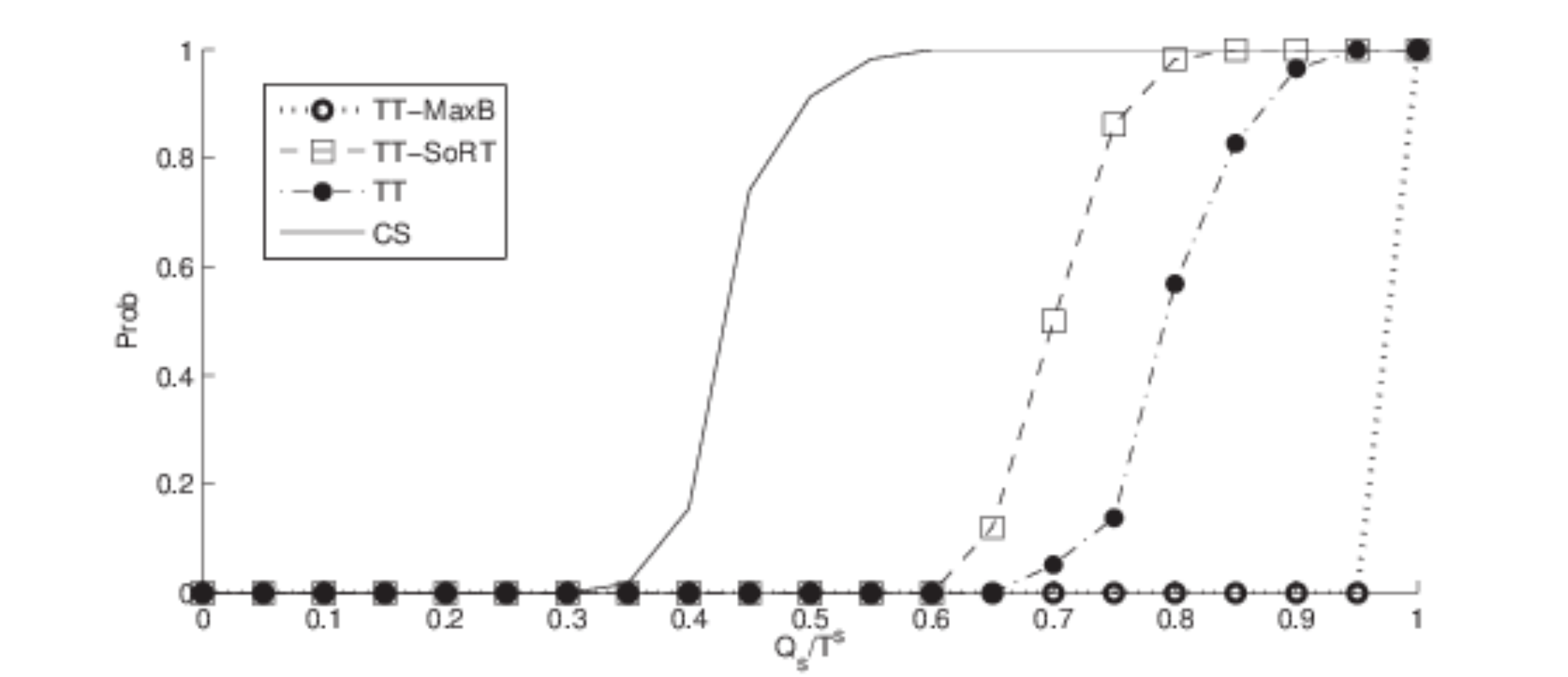}
\caption{Percentage of systems stabilised with different model of
  comptuation vs bandwidth used}
\label{fig:number-stabilised}
\end{figure}
This apparently trivial mathematical issues has profound and far
reaching implication from the practical point of view.  Consider
Figure~\ref{fig:number-stabilised} taken from~\cite{FontanelliPA13}.
We considered 60 randomly generated dynamical systems. For each of
them we synthesised an LQG controller with the same characteristics.
The controller was supposedly implemented by a real-time task having
worst case utilisation $1$ (i.e., maximum computation time equal to
the theoretical period used for the synthesis of the controller).  The
computation time of the task was a randmom variable distributed with a
beta distribution. The plot reports the bandwidth used against the
number of systems that could be stabilised with various models of
computation.  The line TT refers to the hard real time time triggered
model, TT-MaxB refers to a time-triggered model with job drop-out (see
Section~\ref{sec:first-poss-job}), TT-SoRT refers to soft real-time
system with delays (Section~\ref{sec:better-alternative}) and finally
line CS refers to the continuous stream
(Section~\ref{sec:cstream}). As we can see the continuous stream
stabilises most of the systems with 40\% of the bandwidth, with a
substantial saving over the bandwidth required by other models of
computation.


\section{Conclusions}
\label{sec:concl}

In this paper we presented an overview of the state of the art in soft
real-time system. We have presented models for dealing with
uncertainty in execution times and arrival times of tasks, and in
particular we have discussed the resource reservation framework and
the Constant Bandwidth Server algorithm for providing temporal
isolation. We have analysed the typical requirements of soft real-time
tasks, and we have presented a meaningful application to control
system design.

In particular, the latter is important because control systems are
often considered as hard real-time, and subject to very conservative
design methodologies. We have shown how it is possible to improve
their performance by using soft real-time techniques without
renouncing to certification of the properties.

For further information on soft real-time scheduling techniques, the
interested reader can refer to \cite{buttazzo2005soft}. Programming
tools and techniques for soft real-time scheduling with
\texttt{SCHED\_DEADLINE} can be found in
\url{https://github.com/scheduler-tools/rt-app} and
\url{http://retis.sssup.it/rts-like/index.html}.

\printbibliography

\end{document}